\documentclass[a4paper,fleqn]{cas-sc}

\usepackage[numbers]{natbib}

\usepackage{float}
\usepackage{algorithm}
\usepackage{algpseudocode}
\usepackage{subcaption}

\def\tsc#1{\csdef{#1}{\textsc{\lowercase{#1}}\xspace}}
\tsc{WGM}
\tsc{QE}
\tsc{EP}
\tsc{PMS}
\tsc{BEC}
\tsc{DE}

\begin{document}
\let\WriteBookmarks\relax
\def\floatpagepagefraction{1}
\def\textpagefraction{.001}

\shorttitle{SOC}

\shortauthors{Salva\~na et~al.}

\title [mode = title]{A Self-Organized Criticality Model of Extreme Events and Cascading Disasters of Hub and Spoke Air Traffic Networks}                   

%
\author[1,2]{Mary Lai O. Salva\~na}[orcid=0000-0003-4868-7713]

\cormark[1]

\ead{marylai.salvana@uconn.edu}

\ead[url]{marylaisalvana.com}

\address[1]{Department of Statistics, University of Connecticut, 215 Glenbrook Rd., Storrs, Connecticut, 06269, USA}

\author[2]{Harold Jay M. Bolingot}
\ead{harrybo@ateneoinnovation.org}

\author[2]{Gregory L. Tangonan}
\ead{goriot@mac.com}

\address[2]{Ateneo Innovation Center, Ateneo de Manila University, Quezon City, Metro Manila, 1108, Philippines}

\cortext[cor1]{Corresponding author}

\begin{abstract}
Critical infrastructure networks—including transportation, power grids, and communication systems—exhibit complex interdependencies that can lead to cascading failures with catastrophic consequences. These cascaded disasters often originate from failures at critical points in the network, where single-node disruptions can propagate rapidly due to structural dependencies and high-impact linkages. Such vulnerabilities are exacerbated in systems that have been highly optimized for efficiency or have self-organized into fragile configurations over time. The air transportation system in the United States, built on a hub-and-spoke model, exemplifies this type of critical infrastructure. Its reliance on a limited number of high-throughput hubs means that even localized disruptions—particularly those triggered by increasingly frequent and extreme weather events—can initiate cascades with nationwide impacts. We introduce a novel application of the theory of Self-Organized Criticality (SOC) to model and analyze cascading failures in such networks. Through a detailed case study of U.S. airline operations, we show how the SOC model captures the power-law distribution of disruptions and the long-tail risk of systemic failures, reflecting the real-world interplay between structural fragility and external climate shocks. Our approach enables quantitative assessment of network vulnerability, identification of critical nodes, and evaluation of proactive intervention strategies for disaster risk reduction. The results demonstrate that the SOC model successfully replicate the observed statistical patterns of disruption sizes—characterized by frequent small events and rare but severe cascading failures—offering a powerful systems-level framework for infrastructure resilience planning and emergency management. The model provides practitioners with actionable insights for anticipating and mitigating systemic risks in complex, interdependent systems.
\end{abstract}



\maketitle

\section{Introduction}

The catastrophic failures that have increasingly plagued modern air transportation networks reveal a fundamental vulnerability in our interconnected world. When Southwest Airlines canceled over 16,000 flights during the December 2022 winter storm \citep{ABCNews2023Southwest}, when the Federal Aviation Administration's (FAA) Notice to Airmen (NOTAM) system failure grounded all domestic flights in January 2023 \citep{CNN2023FAA}, and when Hurricane Ian's impact cascaded through the entire national aviation network despite affecting only a handful of airports \citep{WashingtonPost2022Ian}, they exposed a troubling reality: the air transportation infrastructure has evolved into a system poised on the edge of chaos \citep{perrow1984normal,comfort1999complexity}.

The hub-and-spoke architecture that defines modern aviation, while remarkably efficient under normal conditions, has produced a network structure exhibiting hallmarks of self-organized criticality (SOC)—a regime where small perturbations can trigger avalanche-like cascading failures \citep{bak1987self}. This architecture concentrates the vast majority of air traffic through a limited number of high-throughput hubs such as ATL (Atlanta), ORD (Chicago O’Hare), DFW (Dallas–Fort Worth), and DEN (Denver). Each of these hubs processes thousands of connecting flights daily, creating a tightly coupled system in which the failure of a single hub can rapidly propagate disruptions across the entire network with devastating efficiency.

\begin{figure}[tb!]
    \centering
    \includegraphics[width=0.6\textwidth]{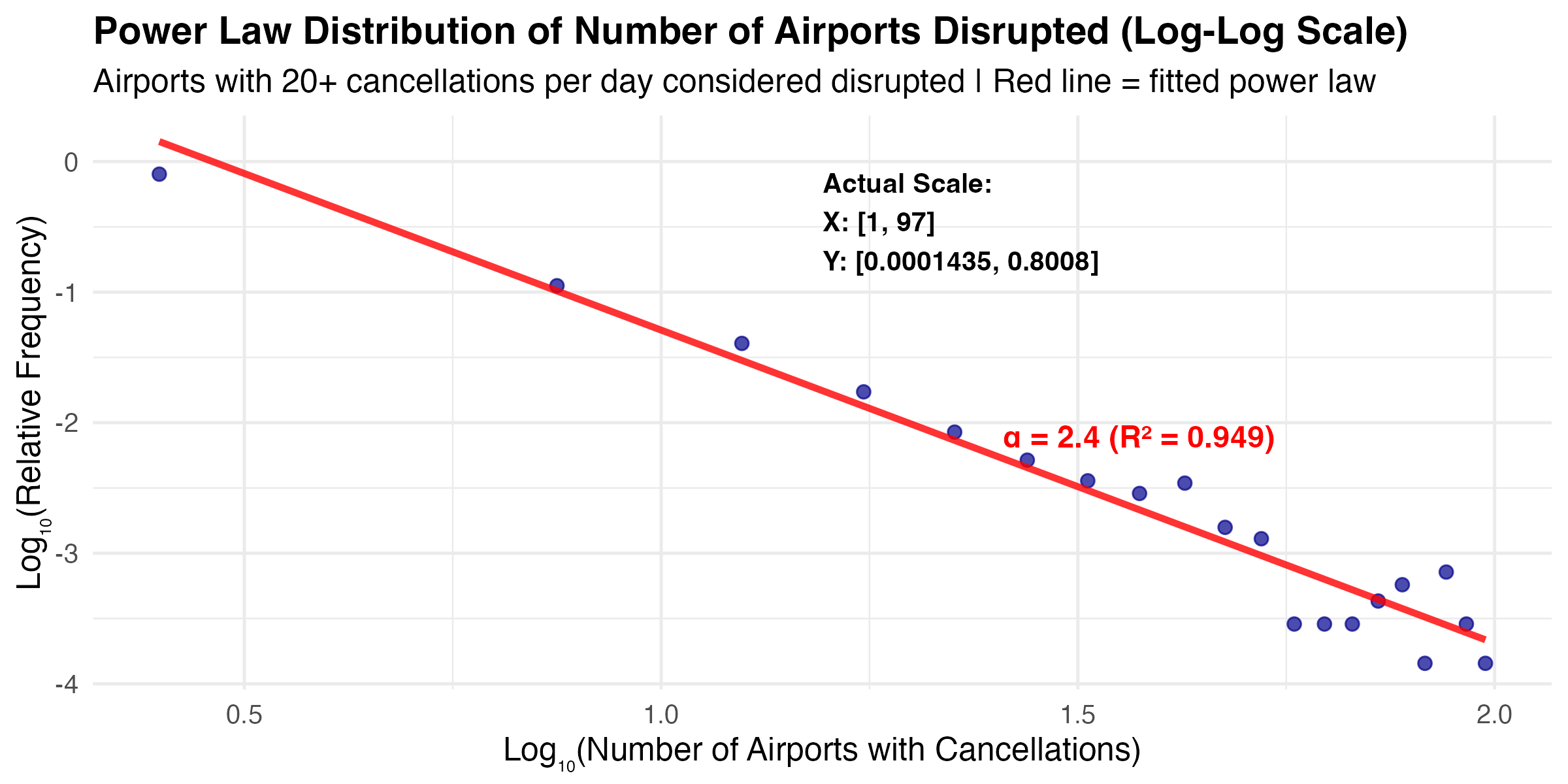}
    \caption{Power-law distribution of airport disruptions in the U.S. air traffic network, exhibiting characteristics of self-organized criticality (SOC). The log-log plot shows a power-law exponent of $\alpha = 2.4$ with a high goodness-of-fit ($R^2 = 0.949$), indicating that while small disruptions are common, large system-wide failures are rare but statistically expected—consistent with the heavy-tailed behavior typical of SOC systems.}
\label{fig:airport_disruption_power_law}
\end{figure}

\begin{figure}[tb!]
    \centering
    \begin{subfigure}[b]{0.48\textwidth}
        \includegraphics[width=\textwidth]{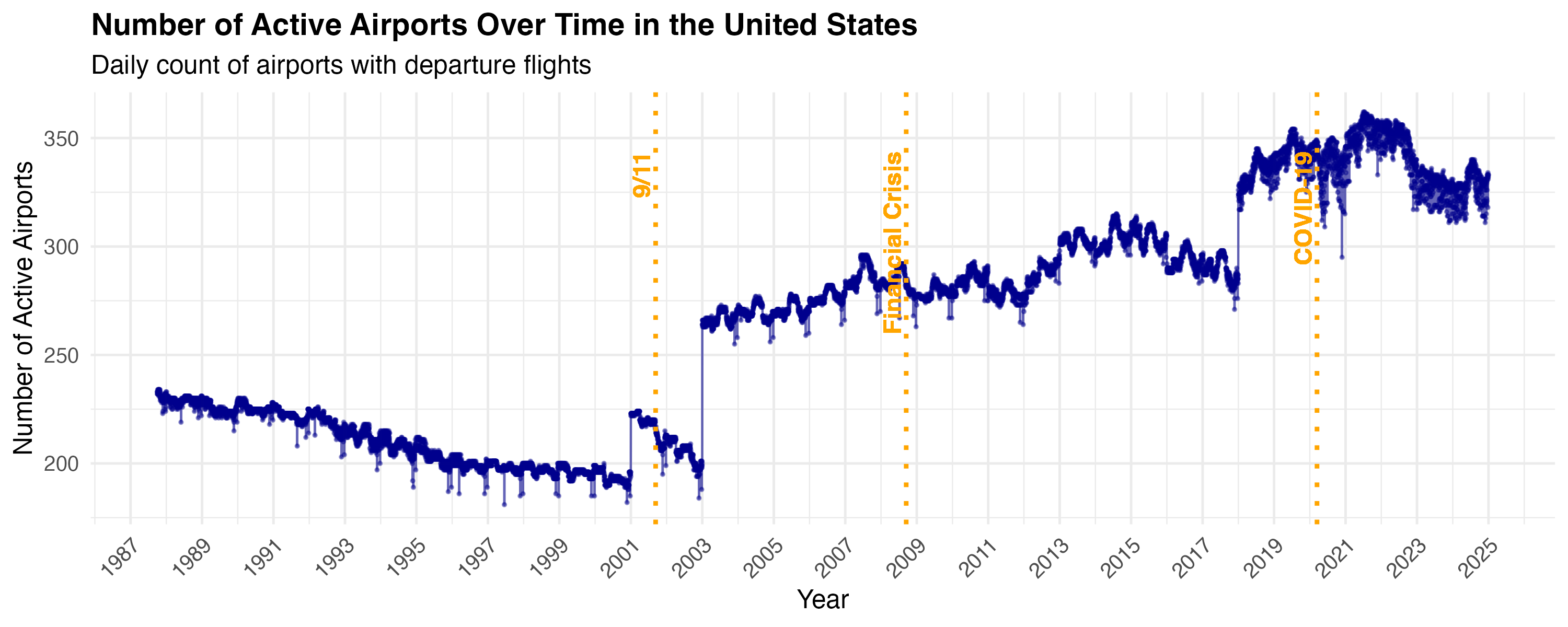}
        \caption{Number of active airports}
        \label{fig:active_airports_sub}
    \end{subfigure}
    \hfill
    \begin{subfigure}[b]{0.48\textwidth}
        \includegraphics[width=\textwidth]{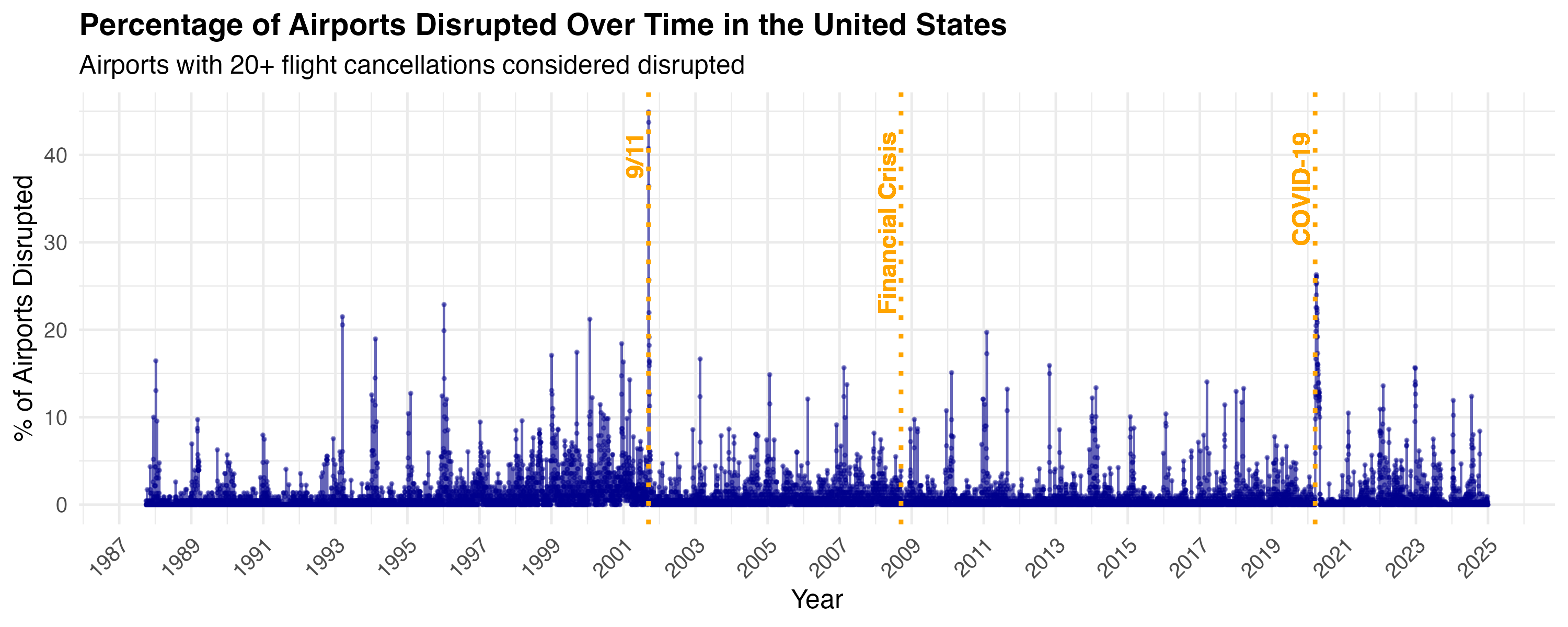}
        \caption{Percentage of airports disrupted}
        \label{fig:airport_disruptions_sub}
    \end{subfigure}
    
    \vspace{0.5cm}
    
    \begin{subfigure}[b]{0.48\textwidth}
        \includegraphics[width=\textwidth]{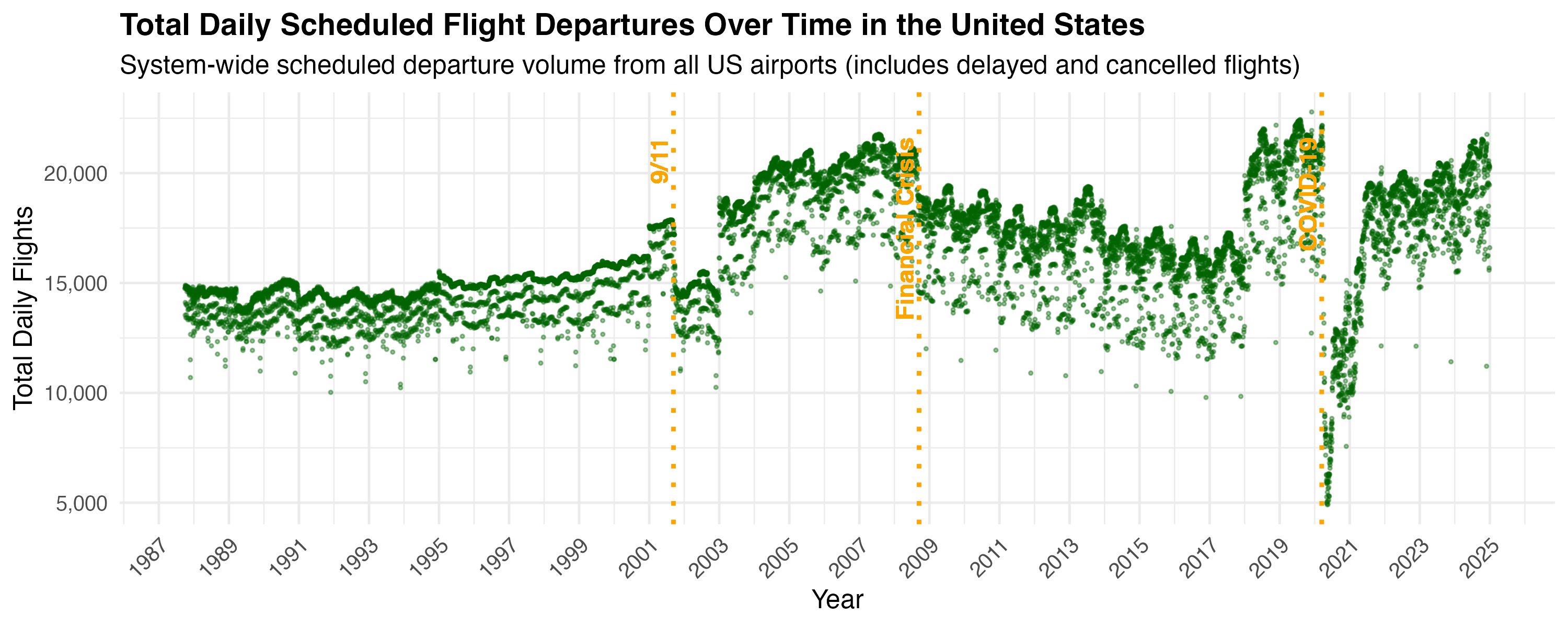}
        \caption{Total daily flight departures}
        \label{fig:flight_departures_sub}
    \end{subfigure}
    \hfill
    \begin{subfigure}[b]{0.48\textwidth}
        \includegraphics[width=\textwidth]{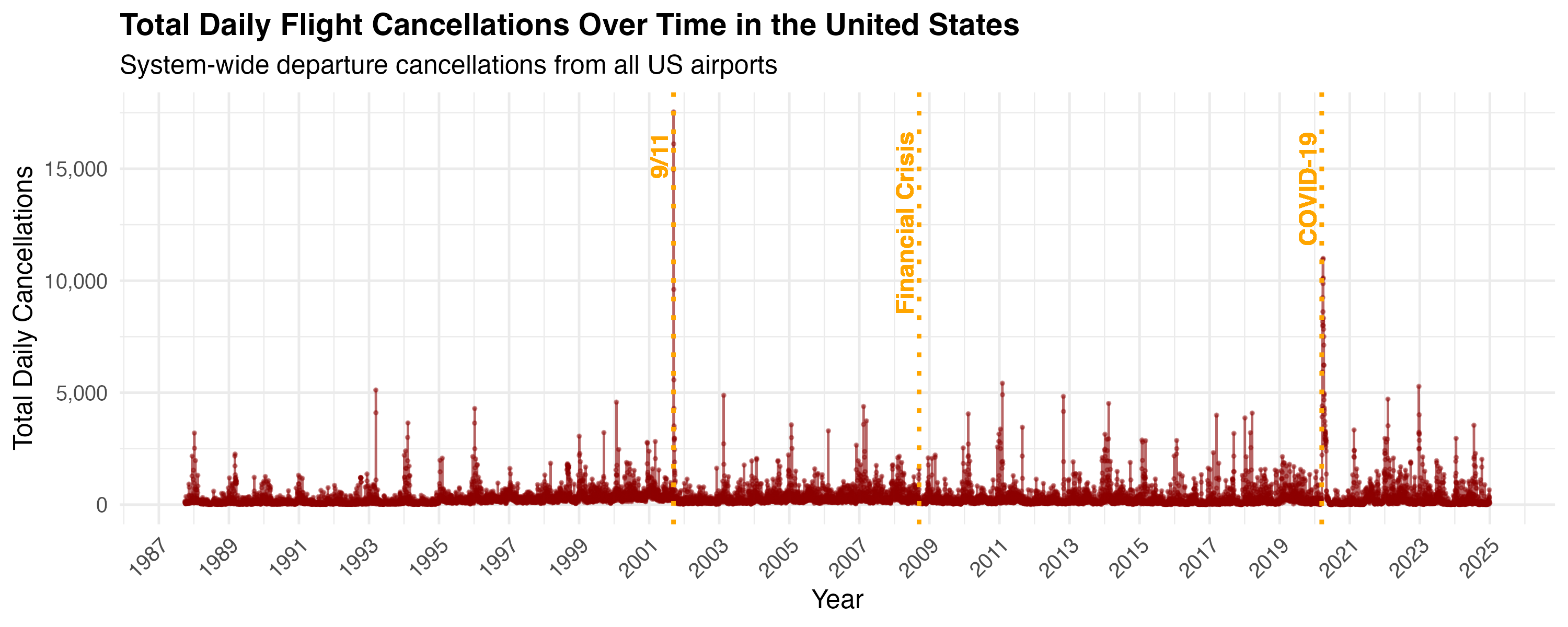}
        \caption{Total daily flight cancellations}
        \label{fig:flight_cancellations_sub}
    \end{subfigure}
    
    \caption{Evolution of the U.S. aviation system from 1987 to 2025 across four key operational metrics. Vertical dotted lines denote major crisis events: 9/11 (2001), the Financial Crisis (2008), and COVID-19 (2020). (a) The number of active airports grows steadily before plateauing, then becomes increasingly sensitive to external shocks. (b) The proportion of airports experiencing disruptions rises over time, reflecting the growing frequency and severity of system-wide failures. (c) Daily flight departures illustrate the expansion of network complexity and traffic volume. (d) Flight cancellations show a rising trend in disruption severity, consistent with the system’s progression toward criticality—where small perturbations can trigger cascading failures.}
    \label{fig:aviation_time_series}
\end{figure}

Our empirical analysis of comprehensive flight operations data from the Bureau of Transportation Statistics \citep{bts2025transtats} shows that the U.S. air traffic network operates in an SOC state. Figure~\ref{fig:airport_disruption_power_law} demonstrates that airport disruptions follow a power-law distribution—a signature of SOC systems—where small disruptions occur frequently while large systemic failures, though rare, emerge as inevitable consequences of the network's structure. This behavior persists across decades of data, as Figure~\ref{fig:aviation_time_series} illustrates by documenting cascading disasters from 1987 to 2025.

The evolution toward criticality reflects the convergence of two reinforcing trends: the relentless optimization of air traffic for business efficiency and the climate-driven intensification of extreme weather. These complementary forces have created a perfect storm of systemic vulnerability—efficiency optimization has eliminated redundancy and resilience buffers precisely as weather shocks have become more frequent and severe. The four panels of Figure~\ref{fig:aviation_time_series} illustrate this transformation: (a) airport network expansion followed by saturation and crisis-driven contraction, (b) disruption patterns exhibiting SOC behavior, (c) operations nearing fundamental capacity limits, and (d) cancellation severity surpassing historical norms. Together, these trends have positioned the aviation system at the edge of chaos—a pattern echoed in power grid failure analysis \citep{salvana2025predicting}—and suggest that cascading failures are now structural features rather than preventable anomalies.

This empirical evidence of SOC behavior in aviation networks reflects a broader pattern of systemic vulnerability emerging across complex systems. Cascading disasters—where initial disruptions propagate through interdependent networks—have drawn increasing attention for their disproportionate and unexpected impacts. In contrast to compound events involving multiple independent hazards, cascading failures originate in tightly coupled, efficiency-optimized systems, where local perturbations rapidly escalate across sectors~\citep{alexander2018magnitude, pescaroli2015cascading, pescaroli2018understanding}. This systems perspective has been applied to multi-hazard interactions, which reveal latent vulnerabilities often overlooked in single-event analyses.~\citep{gill2016hazard, cutter2018compound, deruiter2020we, lawrence2020cascading, ilalokhoin2023model, pescaroli2024progressing}. These dynamics are further amplified by institutional fragilities and coordination gaps~\citep{pescaroli2018understanding, cuartas2021application, huang2024coordination, leppold2025recovery}.

Climate change intensifies these risks by increasing the frequency and severity of extreme events, while aging infrastructure, brittle supply chains, and algorithmic complexity push systems closer to failure~\citep{gratton2022reviewing, little2012managing, verschuur2023systemic, guo2024supply}. Compound extremes—such as wildfire–heatwave–drought linkages—have highlighted the inadequacy of scenario-based planning and underscored the need for systemic models that can capture emergent, network-wide failure patterns~\citep{sutanto2020heatwaves, aghakouchak2018natural, aghakouchak2020climate, barquet2024conceptualising, hoff2025cascading, shimizu2015interconnected}. Yet dominant modeling frameworks remain static and rule-based, lacking the capacity to simulate the nonlinear, evolving dynamics that unfold across space and time~\citep{gong2020cascading, muhlhofer2023generalized, tang2023developing, chen2024modeling, gordan2024protecting, diakakis2025cascade, liu2025risk}.

In contrast, SOC dynamics are emergent: large-scale failures can arise not from a single extreme event but from the gradual accumulation and redistribution of stress across interconnected components. For example, a bridge may appear structurally sound under normal traffic but collapse after repeated minor stresses exceed a threshold. Because tipping points depend on evolving system states—not just observable shocks—models based solely on past events may overlook future vulnerabilities. This limitation is especially acute in dense, interconnected environments, where complexity and fragility grow in tandem~\citep{brunner2024understanding, lee2024reclassifying, kumasaki2016anatomy, huggins2020infrastructural, li2025study, purwar2024qualitative}.

This paper addresses these challenges by introducing a novel application of SOC theory to model cascading failures as emergent outcomes of complex system dynamics. Rather than viewing large-scale disruptions as isolated anomalies, SOC frames them as inevitable outcomes of stress accumulation and structural fragility—offering a generative, systems-level framework for anticipating, diagnosing, and mitigating cascading failures in critical infrastructure networks. The rest of the paper is organized as follows: Section~\ref{sec:soc_model} reviews the theoretical foundations of SOC and its application to complex networks. Section~\ref{sec:methodology} presents the SOC model for the U.S. air traffic network. Section~\ref{sec:simulation} reports simulation experiments demonstrating how structural and operational factors shape cascading behavior. Section~\ref{sec:ews} outlines the potential of SOC modeling for early warning systems. Section~\ref{sec:climate_change} applies SOC to climate stress-testing scenarios. Section~\ref{sec:discussion} concludes the paper and identifies future research directions for critical infrastructure resilience.

\section{Theoretical Foundations of Self-Organized Criticality} \label{sec:soc_model}

SOC represents a major theoretical breakthrough in complex systems science, offering a unified framework for understanding how large-scale catastrophic events can emerge naturally from the intrinsic dynamics of interconnected systems. Unlike traditional approaches that treat major disruptions as exogenous shocks or rare anomalies, SOC theory reveals that catastrophic failures are often structural features of systems operating near critical thresholds. This perspective shifts the focus of infrastructure resilience from preventing isolated failures to managing the dynamics of criticality itself—particularly in systems optimized for efficiency and operating under constant stress.

\subsection{The Revolutionary Insight: Systems on the Edge of Chaos}

The theoretical foundation of SOC emerged from the seminal work of \citet{bak1987self}, who showed that complex systems can spontaneously evolve toward critical states without external tuning. The canonical sandpile model illustrates this: grains of sand accumulate until a critical slope is reached, after which a single grain can trigger avalanches of all sizes, governed by a balance between slow buildup and fast relaxation. SOC systems exhibit three defining characteristics: scale invariance, where statistical patterns persist across scales; universality, where outcomes are independent of micro-level details; and long-range correlations, where local perturbations can affect distant parts of the system. These features produce power-law distributions of event sizes, offering a theoretical basis for why catastrophic failures occur with statistical regularity across domains such as evolution \citep{adami1995self}, solar flares \citep{aschwanden201625}, earthquakes \citep{bhattacharya2007self, chen1991self, sornette1989self}, landslides \citep{hergarten1998self}, financial markets \citep{biondo2015modeling, bouchaud2024self}, economic cycles \citep{scheinkman1994self}, neural dynamics \citep{hesse2014self, plenz2021self}, psychological phenomena \citep{ramos2011self}, ecosystems \citep{levin2005self}, wildfires \citep{gang2022sand}, and power grids \citep{carreras2004evidence, dobson2007complex}.

\subsection{The Mathematics of Criticality: Power-Law Distribution}

The mathematical foundation of SOC is its scale-invariant behavior, expressed through a power-law distribution: \( P(s) \sim s^{-\alpha} \), where \( s \) is the event size and \( \alpha \) is the critical exponent. This distribution implies the absence of a typical event size—unlike normal distributions centered on averages, power laws exhibit heavy tails, where extreme events are far more probable than intuition suggests. The value of \( \alpha \) governs the system’s risk profile: smaller exponents increase the likelihood of large cascades, while larger exponents favor smaller, more frequent events. Because correlations persist across all scales, disruptions can spread broadly in space and time, explaining the heightened sensitivity of air traffic networks as they approach capacity-driven operating limits.

\subsection{SOC in Complex Systems: Universal Critical Behavior}

SOC theory provides a powerful explanation for cascading failures across complex systems. Power grids, internet networks, and financial markets have all been shown to exhibit power-law-distributed failures and nonlinear propagation dynamics. The 2003 Northeast blackout exemplifies this behavior: a line failure in Ohio cascaded across regional networks, ultimately affecting over 50 million people and resulting in an estimated \$10 billion in damages~\citep{cnn2003blackout, nyiso2003blackout}. More recently, \citet{salvana2025predicting} demonstrated that monitoring changes in the critical exponent \( \alpha \) could forecast the 2021 Texas power crisis 6--12 months in advance, with supercriticality behavior (\( \alpha < 1 \)) serving as an early warning indicator.

Building on this foundation, we present a novel framework that models critical infrastructures as SOC systems. This marks a paradigm shift from conventional risk assessments—focused on isolated failure modes—toward a systems-level view where criticality is an emergent property of the system itself. The SOC approach enables quantitative assessment of systemic vulnerability, identification of structurally fragile nodes, and evaluation of intervention strategies. Air traffic networks exemplify these dynamics: hub-and-spoke architectures concentrate stress at central nodes, making them natural cascade triggers. SOC theory suggests that catastrophic failures are not isolated anomalies but statistically expected outcomes driven by network topology, load conditions, and proximity to criticality—rather than by specific initiating events.

\section{The Self-Organized Criticality Model of the U.S. Air Traffic Network} \label{sec:methodology}

\begin{figure}[tb!]
\centering
\includegraphics[width=0.7\textwidth]{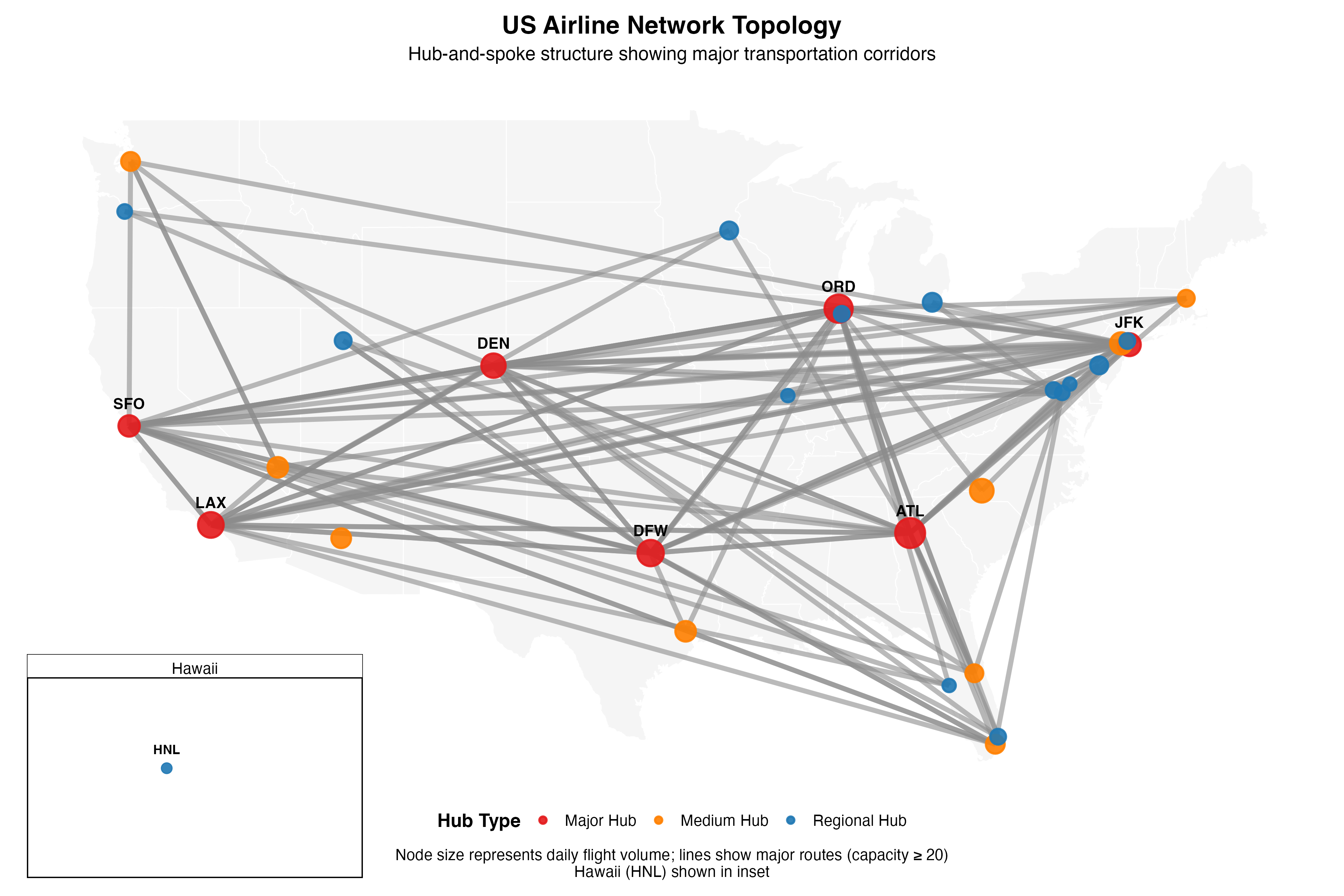}
\caption{Topology of the U.S. airline network, highlighting its hub-and-spoke structure. Major hubs (red), medium hubs (orange), and regional hubs (blue) are distinguished by node color; node size reflects hub strength, and edge thickness indicates route capacity. The network exhibits scale-free properties with pronounced clustering around major airports. All airport identifiers follow standard International Air Transport Association (IATA) three-letter codes (e.g., ATL for Atlanta, ORD for Chicago O'Hare).}
\label{fig:network_topology}
\end{figure}

Our modeling framework formalizes the principle that any SOC system must incorporate five core elements that collectively determine the system’s critical behavior and the statistical properties of cascading failures:

\begin{enumerate}
   \item \textbf{Network Configuration}: Defines the pathways through which failures propagate, forming the topological substrate on which critical phenomena emerge. Figure~\ref{fig:network_topology} depicts the hierarchical hub-and-spoke structure of U.S. aviation. Examples include:

   \begin{itemize}
       \item Scale-free networks with hub-and-spoke architecture, where traffic concentrates at major hubs like ATL
       \item Small-world networks, such as power grids, with local clustering and long-range connections
       \item Regular lattice structures modeling wildfire spread through nearest-neighbor interactions
       \item Modular networks representing urban infrastructure systems—such as water, transportation, and communication—where tightly coupled subsystems interact through sparse interlinks
   \end{itemize}
   
   \item \textbf{Stress Accumulation Rule}: Describes how operational pressure builds up in the system over time, reflecting the slow progression toward criticality seen in real infrastructure. In aviation, stress arises from operational bottlenecks, environmental hazards, and workforce dynamics. Examples include:
   
   \begin{itemize}
       \item Uniform random accumulation simulating evenly distributed passenger demand
       \item Preferential buildup at high-traffic hubs due to dense connecting flight schedules
       \item Spatially correlated stress representing weather systems affecting multiple nodes simultaneously
       \item Time-varying accumulation reflecting seasonal travel surges (e.g., holiday peaks at MCO and LAS)
   \end{itemize}
   
   \item \textbf{Failure Condition}: Specifies the threshold at which individual nodes become unstable and trigger redistributive events, capturing capacity constraints. Examples include:
   
   \begin{itemize}
       \item Fixed thresholds modeling standardized runway capacity across airports
       \item Degree-dependent thresholds where large hubs withstand higher stress than regional nodes
       \item Stochastic thresholds incorporating weather-induced capacity variation
       \item Dynamic thresholds that adjust under crew shortages or mechanical failures
   \end{itemize}
   
   \item \textbf{Stress Redistribution Rule}: Determines how stress released by failed nodes transfers to neighboring nodes, governing how local failures cascade. Examples include:
   
   \begin{itemize}
       \item Equal redistribution across all connected neighbors
       \item Weighted redistribution where high-capacity nodes absorb proportionally more stress
       \item Distance-decaying redistribution favoring nearby airports
       \item Selective redistribution to least-loaded neighbors to mitigate cascade amplification (e.g., stress from failed EWR preferentially shifts to underutilized PHL rather than already-burdened JFK)
   \end{itemize}

   \item \textbf{Cascade Propagation}: Describes how failures recursively trigger other failures until the system reaches a stable state. This element captures the emergent nature of large-scale disruptions and distinguishes SOC models from isolated failure simulations. Examples include:

   \begin{itemize}
       \item Recursive toppling where newly overloaded nodes fail in subsequent time steps
       \item Avalanche termination criteria (e.g., all node loads fall below thresholds)
       \item Propagation-limited cascades constrained by topology or spatial buffers
       \item Multi-round propagation with partial recovery between steps to model real-time mitigation
   \end{itemize}
\end{enumerate}

These five elements jointly shape the system’s emergent behavior, including the tail properties of cascading failure distributions. This study presents the first systematic application of SOC principles to the analysis of air traffic networks. It shows that extreme disruptions—the frequency and severity of large-scale failures—emerge not solely from external shocks, but from the local interaction rules embedded within the system. In the context of aviation, this perspective reveals how operational policies, network architectures, and redistribution protocols interact to shape systemic vulnerability, offering a scientific foundation for evidence-based strategies to enhance infrastructure resilience.

\begin{figure}[tb!]
   \centering
   \includegraphics[width=\textwidth]{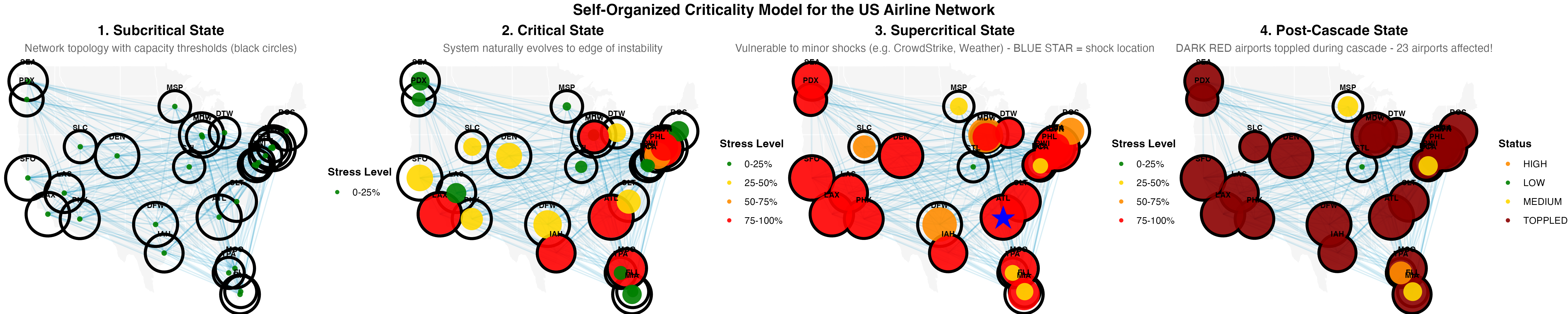}
   \caption{Progression of the U.S. airline network through four phases of SOC: (1) Subcritical State—low system-wide stress with all nodes operating below capacity (green); (2) Critical State—heterogeneous stress accumulation as the system approaches instability (yellow to red); (3) Supercritical State—cascading failures initiated by a local shock (blue star); and (4) Post-Cascade State—widespread collapse with toppled airports (dark red) and partial recovery. Node colors represent stress levels (green to red), black rings denote capacity thresholds, and node size reflects relative importance.}
   \label{fig:soc_evolution}
\end{figure}

Our SOC model captures the characteristic evolution of critical infrastructure systems as they transition through distinct phases of fragility. Figure~\ref{fig:soc_evolution} illustrates this progression in the U.S. airline network—from a stable subcritical state to a critical state with elevated stress, followed by supercritical conditions where minor shocks can trigger widespread disruptions, and culminating in a post-cascade state where the network reorganizes. These phases emerge naturally from the interplay of the model’s five core elements. The SOC framework reproduces essential dynamics: systems drift toward criticality under normal operations, undergo abrupt cascades when local thresholds are breached, and reconfigure with altered stress distributions. Varying model parameters reveals distinct tail behaviors in the distribution of cascade sizes, highlighting a spectrum of systemic risk scenarios. These results underscore the importance of network design and operational policy in distinguishing routine disruptions from tipping points that can lead to systemic collapse.

\section{Simulation Study} \label{sec:simulation}

We conduct simulation experiments to investigate how structural and operational factors shape SOC behavior and the distribution of cascading failures in airport networks. The baseline configuration includes:

\begin{itemize}
    \item \textit{Failure threshold}: Uniform, $\theta = 1$;
    \item \textit{Stress accumulation}: Random addition of $\delta_s = 0.1$ units per step.
\end{itemize}

We vary the redistribution parameter $\beta \in \{0.5, 0.75, 0.99\}$ to model different stress propagation regimes. Lower values represent highly dissipative systems; higher values approximate near-total redistribution. These policies are tested across three network topologies:

\begin{itemize}
    \item \textit{Hub-and-Spoke}: Centralized around major hubs (e.g., ATL), with regional nodes dependent on hub connectivity;
    \item \textit{Point-to-Point}: Decentralized with dense interconnectivity among secondary airports (e.g., LAS, PHX);
    \item \textit{Fragmented Regional}: Decentralized subregions with strong internal links and limited cross-region connectivity.
\end{itemize}

\begin{figure}[tb!]
\centering
\begin{subfigure}[b]{\textwidth}
   \centering
   \includegraphics[width=0.9\textwidth]{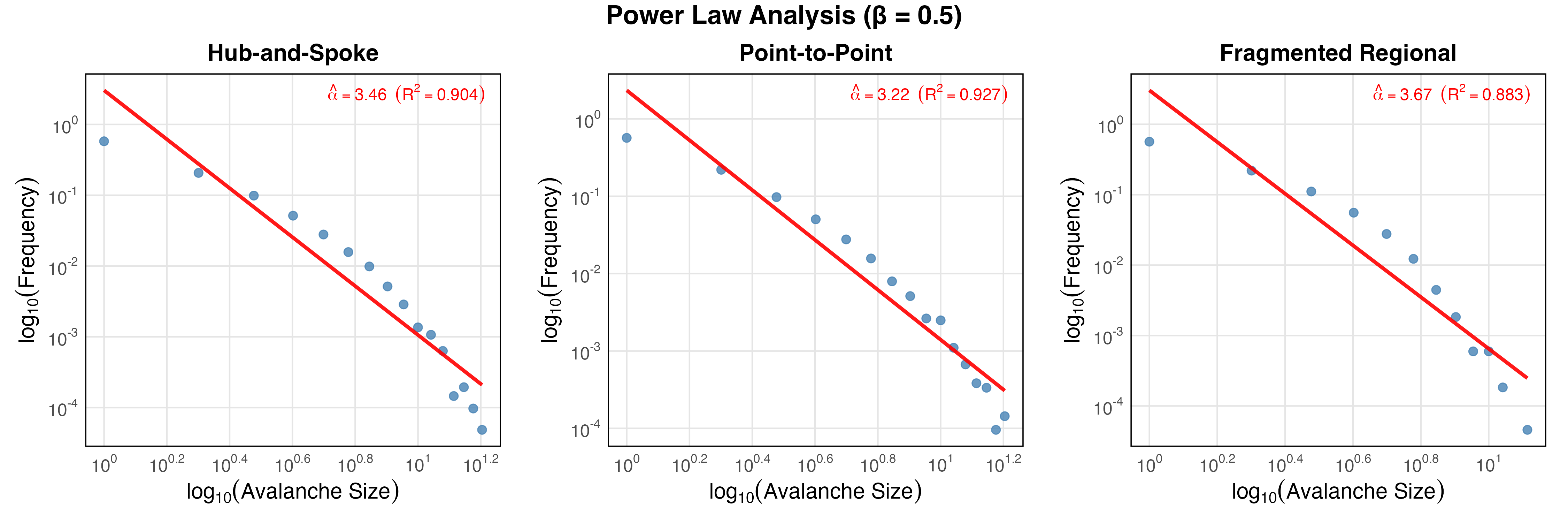}
   \caption{Low Redistribution}
   \label{fig:network_beta_0.5}
\end{subfigure}
\vspace{0.5cm}
\begin{subfigure}[b]{\textwidth}
   \centering
   \includegraphics[width=0.9\textwidth]{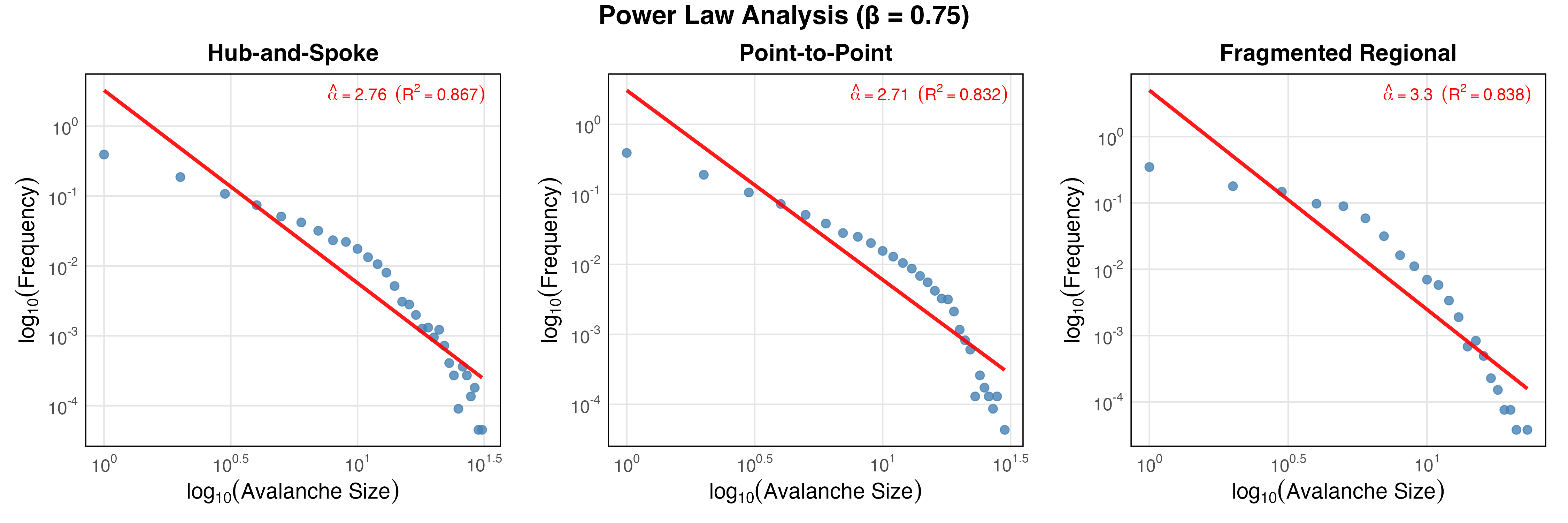}
   \caption{Moderate Redistribution}
   \label{fig:network_beta_0.75}
\end{subfigure}
\vspace{0.5cm}
\begin{subfigure}[b]{\textwidth}
   \centering
   \includegraphics[width=0.9\textwidth]{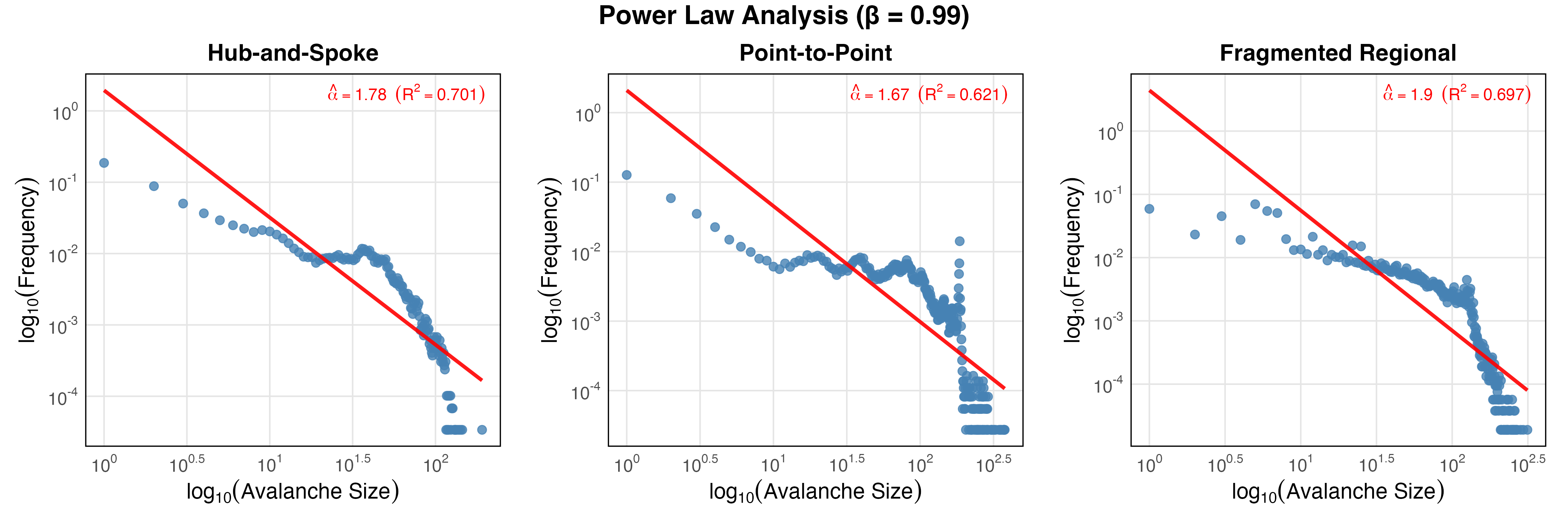}
   \caption{High Redistribution}
   \label{fig:network_beta_0.9}
\end{subfigure}
\caption{Power-law distributions of avalanche sizes across three network configurations under varying stress redistribution parameters ($\beta$). As $\beta$ increases, topological effects become more pronounced, with clearer differentiation in tail behavior and scaling exponents under high redistribution efficiency.}
\label{fig:network_config_comparison}
\end{figure}

\begin{figure}[tb!]
\centering
\begin{subfigure}[b]{\textwidth}
   \centering
   \includegraphics[width=0.8\textwidth]{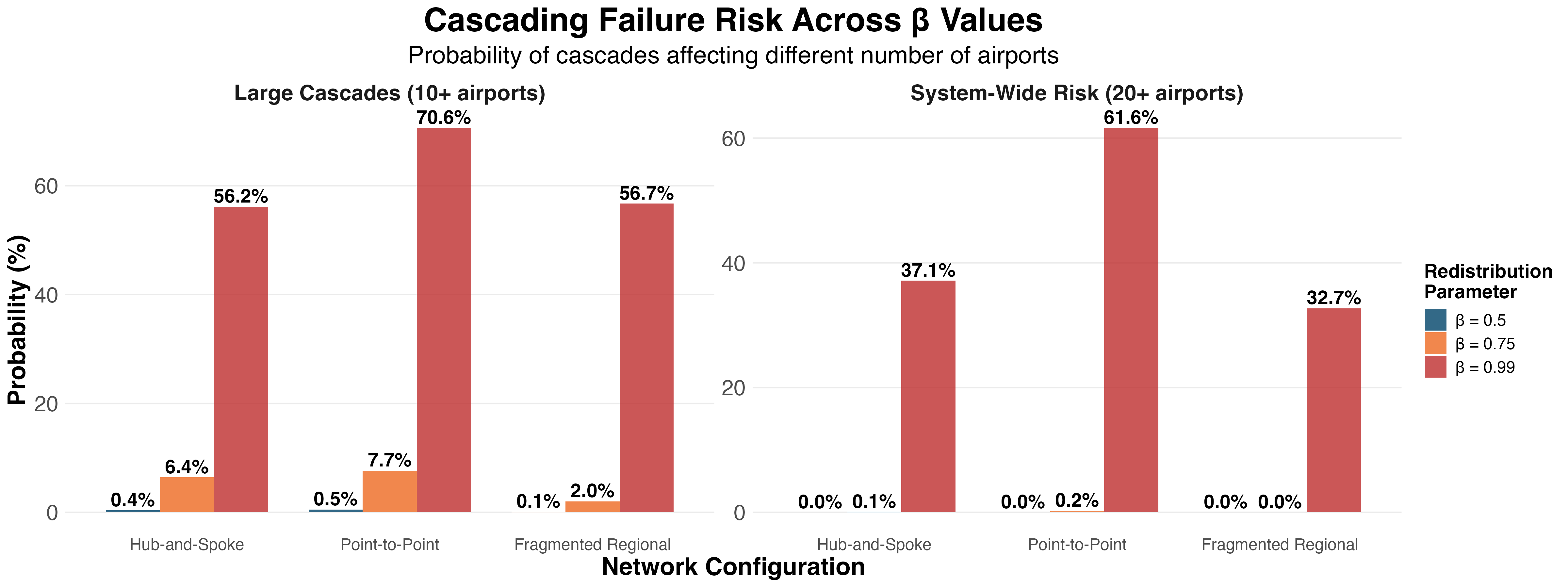}
   \caption{}
   \label{fig:cascading_failure_risk}
\end{subfigure}

\vspace{0.5cm}

\begin{subfigure}[b]{\textwidth}
   \centering
   \includegraphics[width=0.8\textwidth]{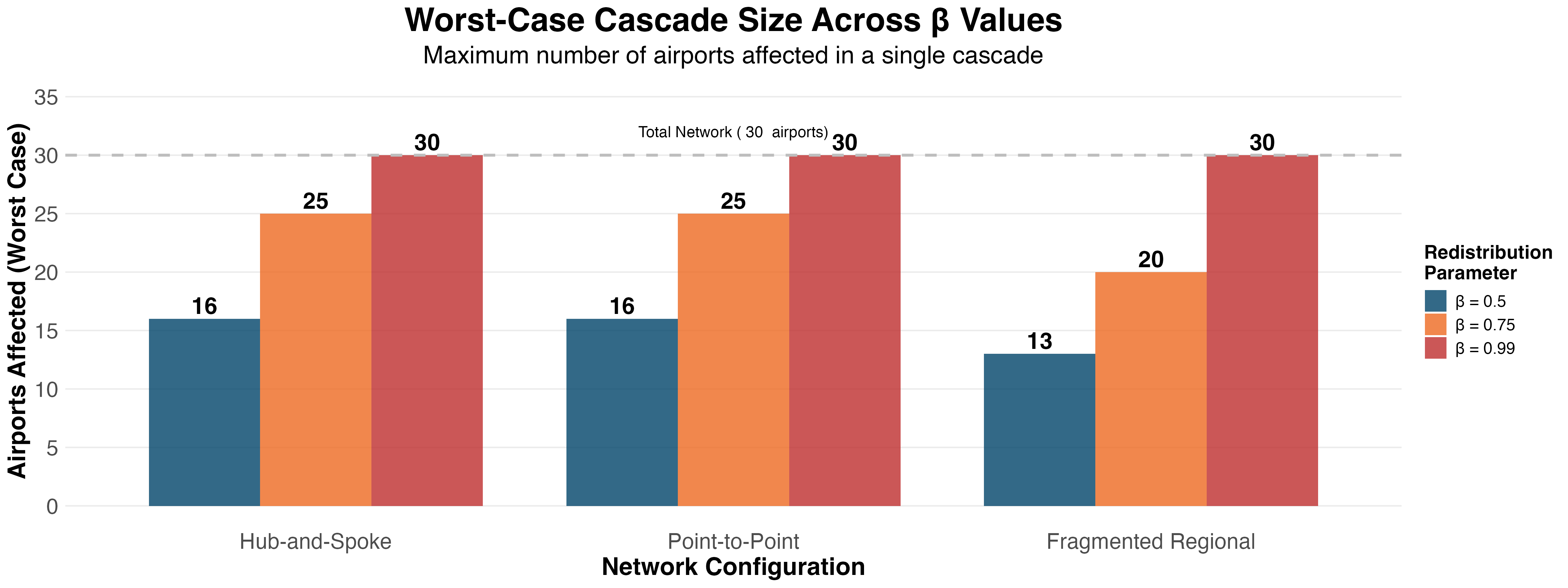}
   \caption{}
   \label{fig:worst_case_impact}
\end{subfigure}

\vspace{0.5cm}

\begin{subfigure}[b]{\textwidth}
   \centering
   \includegraphics[width=0.8\textwidth]{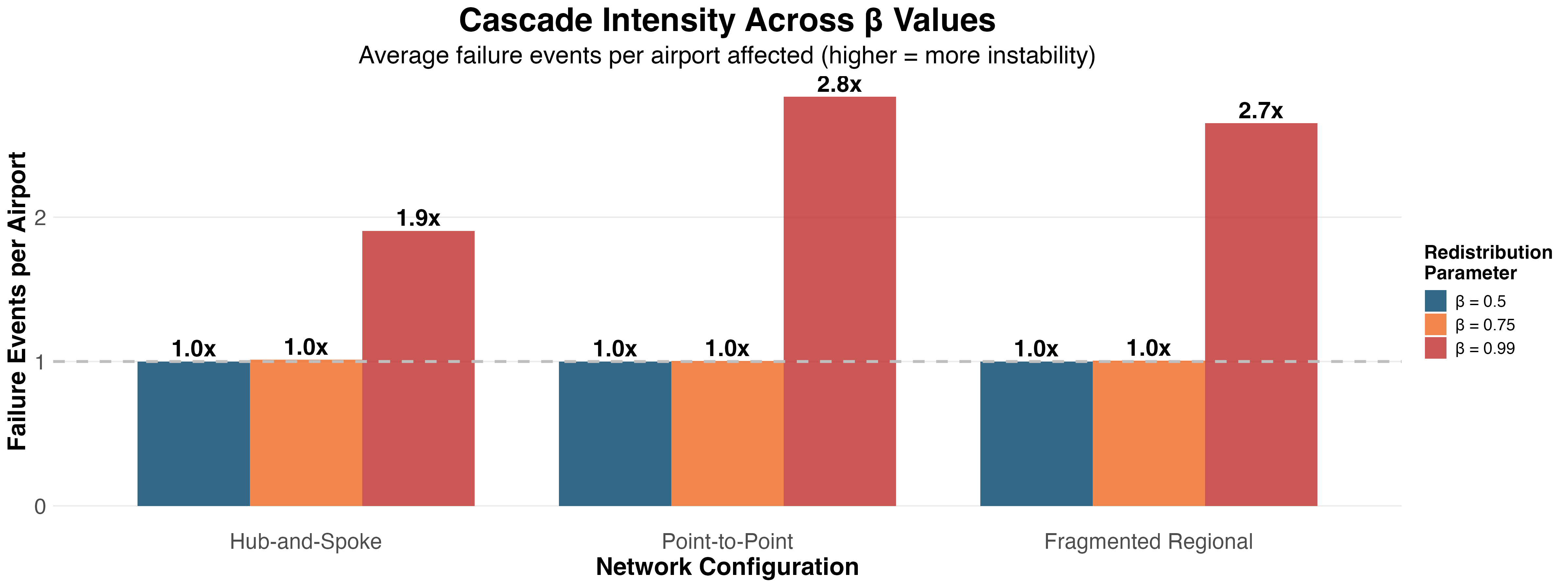}
   \caption{}
   \label{fig:cascade_intensity}
\end{subfigure}
\caption{Cascading failure risk across network configurations, showing the probability of large-scale events, worst-case cascade sizes, and average cascade intensity under varying stress redistribution levels. Results highlight how network topology shapes systemic vulnerability, especially under high-efficiency redistribution.}
\label{fig:network_risk_analysis}
\end{figure}

The simulation results reveal striking differences in cascade behavior across network configurations, with effects becoming increasingly pronounced as the redistribution parameter $\beta$ increases (Figure~\ref{fig:network_config_comparison}). At low redistribution efficiency ($\beta = 0.5$), all networks exhibit similar behavior, characterized by steep power-law exponents ($\hat{\alpha}$ between 3.22 and 3.67) and limited cascade sizes (maximum of 13–16 airports). In this regime, constrained redistribution acts as a natural firebreak, containing failures regardless of topology. As redistribution efficiency increases to moderate levels ($\beta = 0.75$), topological differences begin to emerge. The Point-to-Point network displays a 7.7\% probability of large cascades and a 0.2\% system-wide risk. The Hub-and-Spoke network follows with a 6.4\% probability of large cascades and 0.1\% system-wide risk, while the Fragmented Regional network maintains strong containment, showing only a 2.0\% probability of large cascades and a maximum impact of 20 airports (Figure~\ref{fig:network_risk_analysis}). 

At high redistribution efficiency ($\beta = 0.99$), topology effects become dominant. Both Hub-and-Spoke and Point-to-Point configurations exhibit extreme vulnerability, with large cascade probabilities exceeding 56\% and system-wide risks reaching 37.1\% and 61.6\%, respectively. Even the Fragmented Regional network—despite its decoupled design—shows a 56.7\% probability of large cascades and 32.7\% system-wide risk, indicating that highly efficient stress transfer can breach regional boundaries. These findings reveal a fundamental trade-off in aviation network design: operational improvements that enhance redistribution also increase systemic fragility by enabling failure propagation. The Hub-and-Spoke network is vulnerable due to centralization, while the Point-to-Point network’s dense interconnectivity creates multiple cascade pathways under high $\beta$. By contrast, the Fragmented Regional network exhibits higher resilience in most scenarios—yet remains susceptible under extreme redistribution. This underscores the need for balanced design strategies that weigh efficiency against fragility in the face of cascading failure dynamics.

\section{SOC-Based Early Warning System: Predicting Cascade Propagation Patterns} \label{sec:ews}

Building on the SOC framework, we develop an early warning system (EWS) to identify vulnerable nodes and anticipate cascade propagation before catastrophic failures occur. This provides actionable insights for proactive risk management and resource allocation in critical infrastructures.

\subsection{Early Warning System Framework}

The EWS monitors vulnerability indicators under the most cascade-prone operational scenario and applies five complementary metrics that capture distinct dimensions of risk and cascade dynamics:

\begin{itemize}
\item \textit{Airport Criticality Index}: A composite measure integrating structural and operational factors:
    \begin{equation}
    C_i = w_1 \cdot \frac{k_i}{\max_j(k_j)} + w_2 \cdot \frac{B_i}{\max_j(B_j)} + w_3 \cdot \frac{F_i}{\max_j(F_j)}
    \label{eq:criticality}
    \end{equation}
    where $k_i$ is degree centrality, $B_i$ is betweenness centrality, and $F_i$ is failure frequency. We assign weights $w_1 = 0.3$, $w_2 = 0.3$, and $w_3 = 0.4$, with greater emphasis on observed failure patterns.

\item \textit{Cascade Participation Frequency}: Measures how often airport $i$ is involved in cascade events:
    \begin{equation}
    P_i = \sum_{c=1}^{N_c} I(i \in A_c)
    \label{eq:cascade_participation}
    \end{equation}
    where $N_c$ is the total number of cascades and $A_c$ is the set of airports affected in cascade $c$.

\item \textit{Avalanche Trigger Propensity}: Measures how frequently airport $i$ initiates a cascading failure:
    \begin{equation}
    T_i = \sum_{c=1}^{N_c} I(i = \text{trigger}(c))
    \label{eq:trigger_propensity}
    \end{equation}

\item \textit{Contagion Risk}: Quantifies probability flowing from airport $i$, indicating its influence on secondary failures:
    \begin{equation}
    R_i^{\text{out}} = \sum_{j \neq i} P(j|i)
    \label{eq:contagion_risk}
    \end{equation}
    where $P(j|i) = N_{i \to j}/N_i$ is the conditional probability that $j$ fails given $i$ has failed.

\item \textit{Susceptibility Risk}: Measures how often airport $j$ experiences secondary failures from upstream disruptions:
    \begin{equation}
    R_j^{\text{in}} = \sum_{i \neq j} P(j|i)
    \label{eq:susceptibility_risk}
    \end{equation}
\end{itemize}

\subsection{Results and Validation}

\begin{figure}[tb!]
   \centering
   \includegraphics[width=\textwidth]{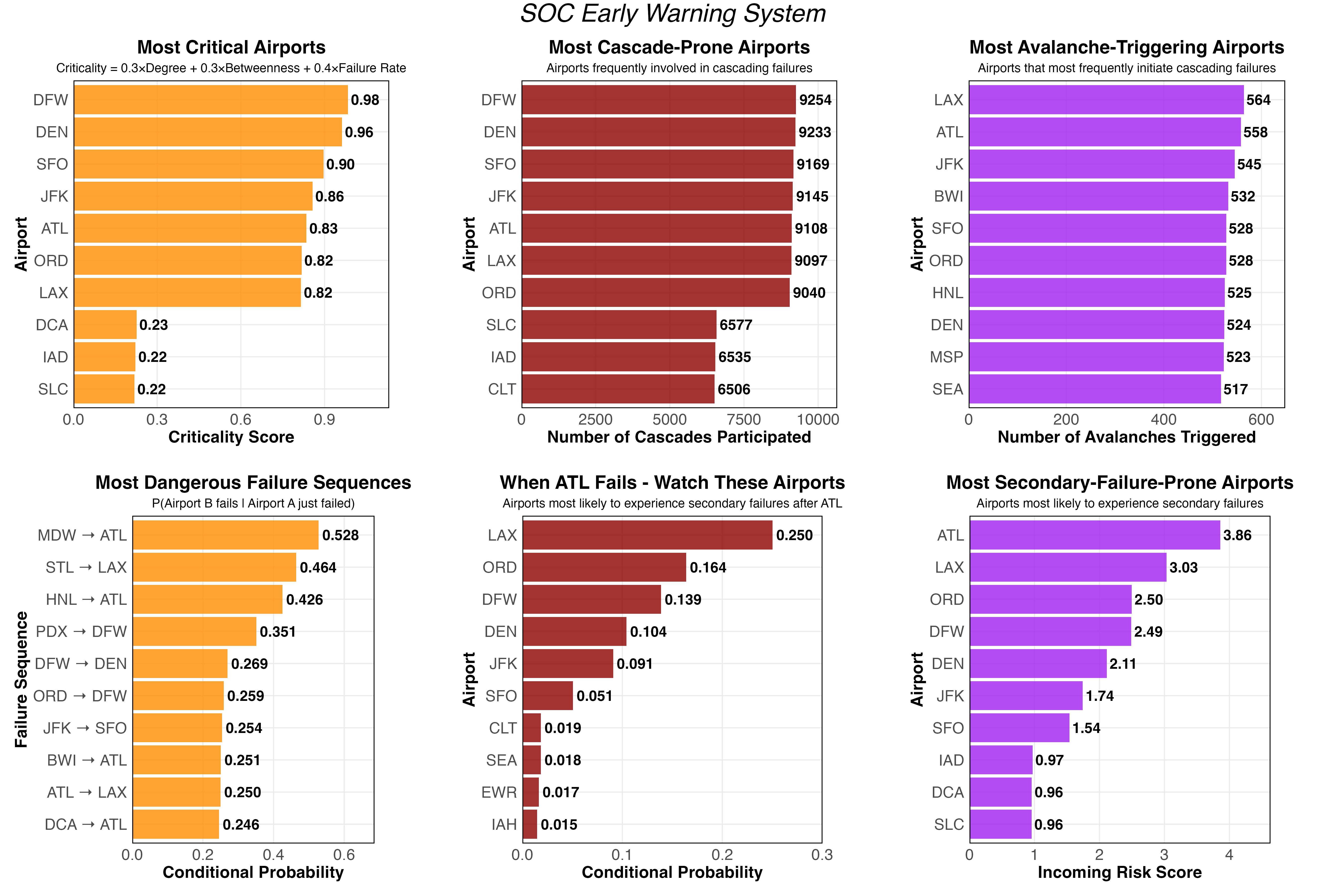}
   \caption{Comprehensive SOC-based EWS analysis. Top panels show static vulnerability metrics: criticality scores (left), cascade participation frequency (middle), and avalanche trigger propensity (right). Bottom panels present dynamic failure propagation analysis: most dangerous failure sequences (left), conditional failure probabilities following ATL failure (middle), and overall vulnerability to secondary failures (right).}
   \label{fig:early_warning}
\end{figure}

Figure~\ref{fig:early_warning} summarizes EWS results, revealing distinct patterns of vulnerability across the network. The analysis uncovers multiple risk dimensions, each requiring targeted intervention strategies.

Static vulnerability assessment ranks major hubs highest across all metrics. DFW leads in criticality (0.98), followed by DEN (0.96) and SFO (0.90), reflecting high connectivity and stress exposure. Cascade participation is also dominated by DFW (9,254), DEN (9,233), and SFO (9,169), while avalanche initiation is most frequent at LAX (564), ATL (558), and JFK (545).

Dynamic propagation analysis identifies the most dangerous sequence as MDW $\rightarrow$ ATL (0.528 conditional probability), followed by STL $\rightarrow$ LAX (0.464) and HNL $\rightarrow$ ATL (0.426). When ATL fails, the most likely secondary failures occur at LAX (0.250), ORD (0.164), and DFW (0.139). System-wide susceptibility is highest for ATL (3.86), LAX (3.03), and ORD (2.50), underscoring their roles as high-risk convergence points.

The convergence of DFW, DEN, and SFO across static metrics, and ATL’s prominence in dynamic propagation pathways, highlights key nodes requiring early intervention. Critical airports benefit from enhanced monitoring and structural reinforcement; cascade-prone airports need stress-buffering mechanisms; and trigger-prone airports merit preemptive stabilization to reduce cascade risk.

These simulation-based results demonstrate the SOC framework’s ability to identify latent vulnerabilities and anticipate propagation pathways. The EWS offers both strategic risk profiling and tactical sequence prediction, enabling more proactive risk governance. It generalizes across network types and operational regimes, making it broadly applicable to aviation and other interdependent infrastructure systems.

\section{SOC-Based Climate Vulnerability Assessment: Hurricane-Driven Cascade Amplification} \label{sec:climate_change}

The SOC framework provides a basis for operational stress testing to assess climate-driven cascade amplification. This enables policy-oriented scenario analysis by systematically varying hurricane intensity \( I_h \) as a stress-testing parameter, allowing exploration of climate trajectories and resilience strategies under increasing environmental pressure.

\begin{figure}[tb!]
\centering
\begin{subfigure}[b]{0.48\textwidth}
   \includegraphics[width=\textwidth]{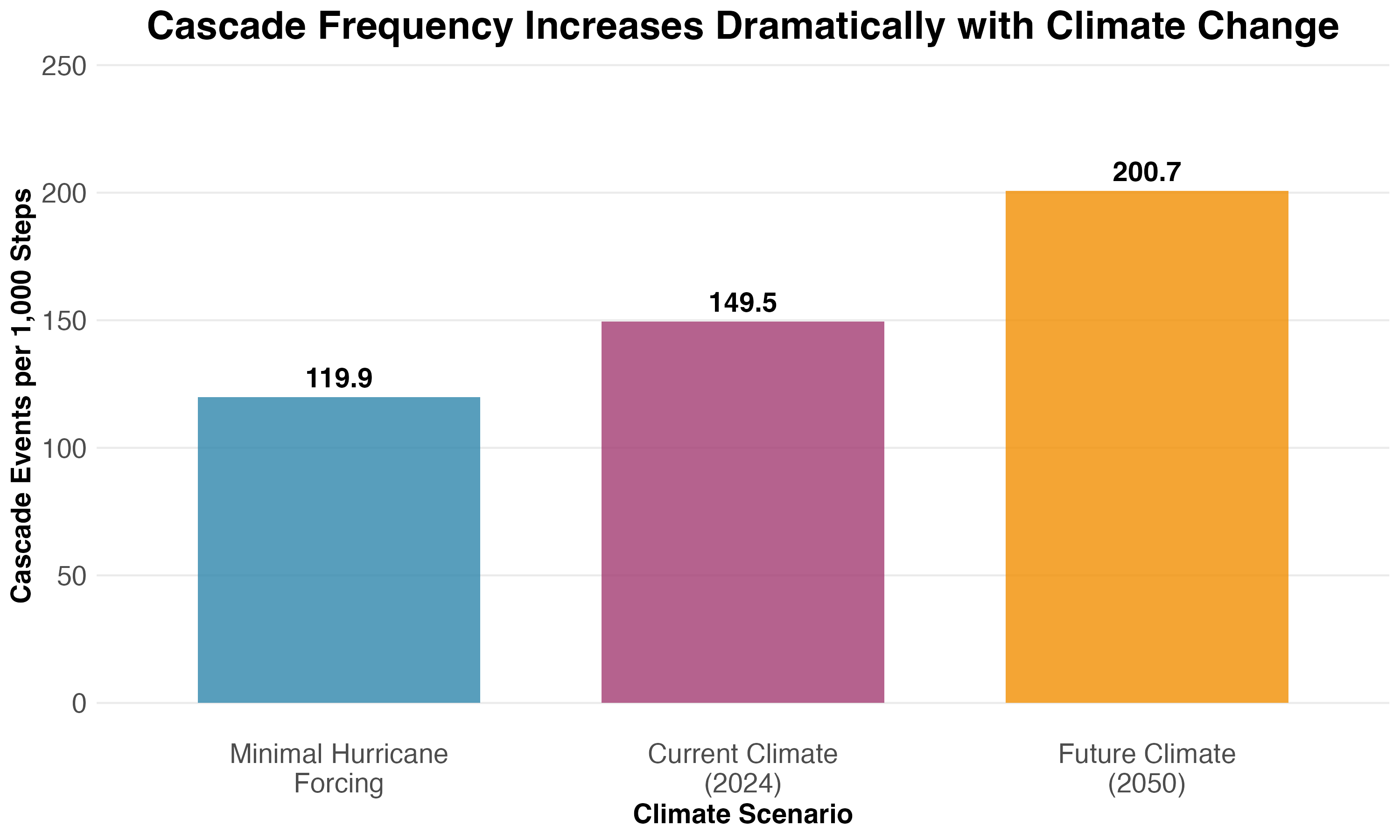}
   \caption{Cascade frequency increases by 67\%.}
   \label{fig:cascade_frequency}
\end{subfigure}
\hfill
\begin{subfigure}[b]{0.48\textwidth}
   \includegraphics[width=\textwidth]{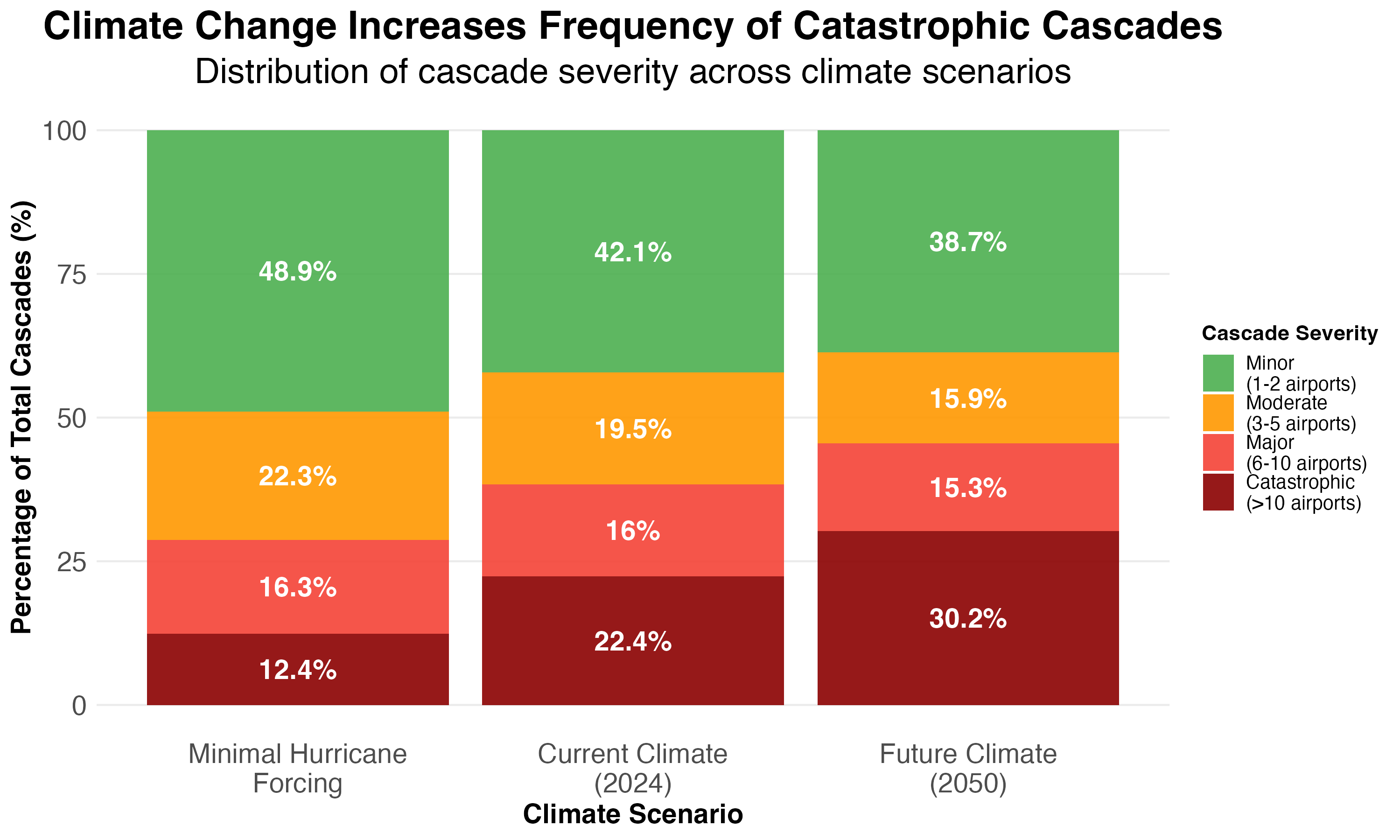}
   \caption{Catastrophic cascades rise from 12.4\% to 30.2\%.}
   \label{fig:cascade_severity}
\end{subfigure}

\vspace{0.5cm}

\begin{subfigure}[b]{0.48\textwidth}
   \includegraphics[width=\textwidth]{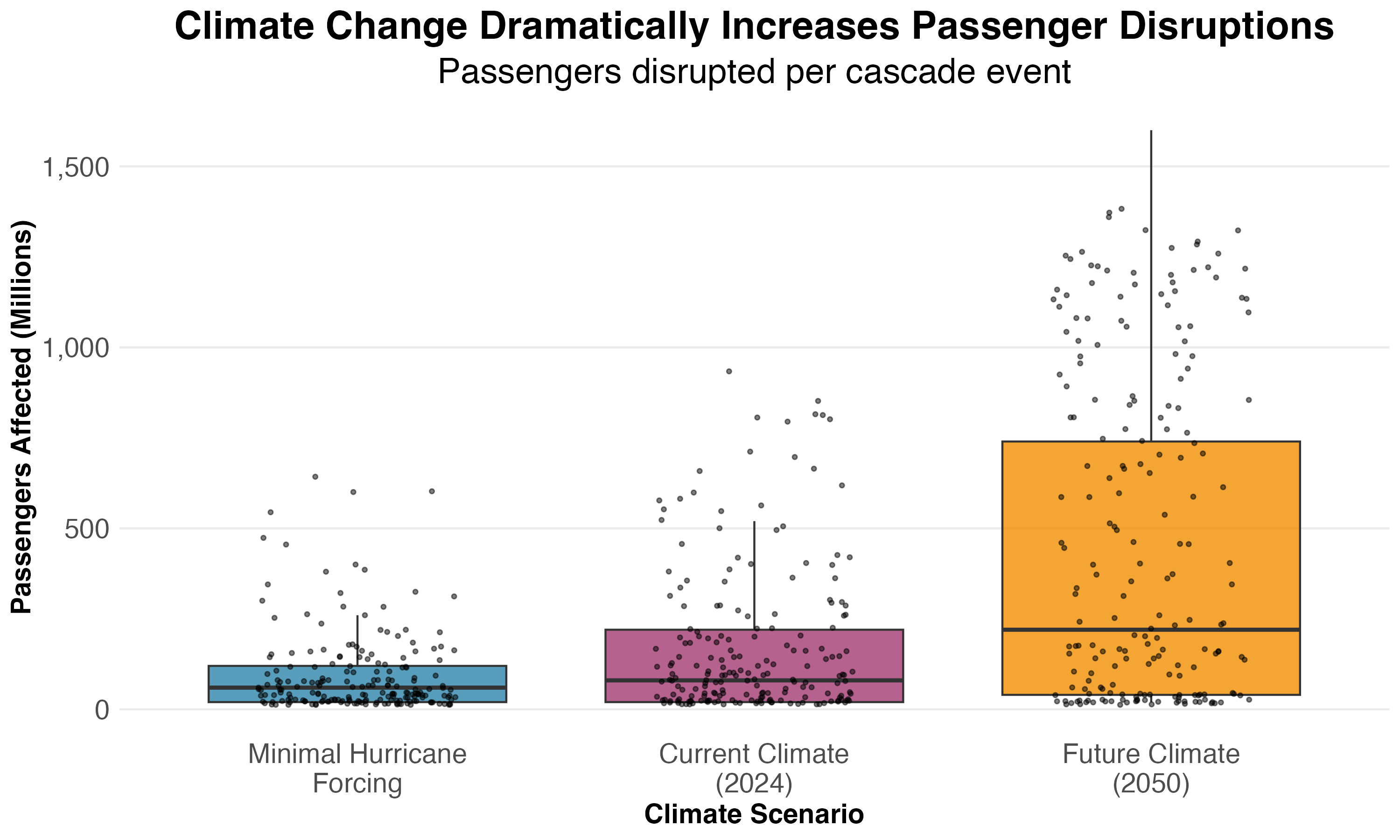}
   \caption{Climate change vastly increases passenger disruptions.}
   \label{fig:passenger_impact}
\end{subfigure}
\hfill
\begin{subfigure}[b]{0.48\textwidth}
   \includegraphics[width=\textwidth]{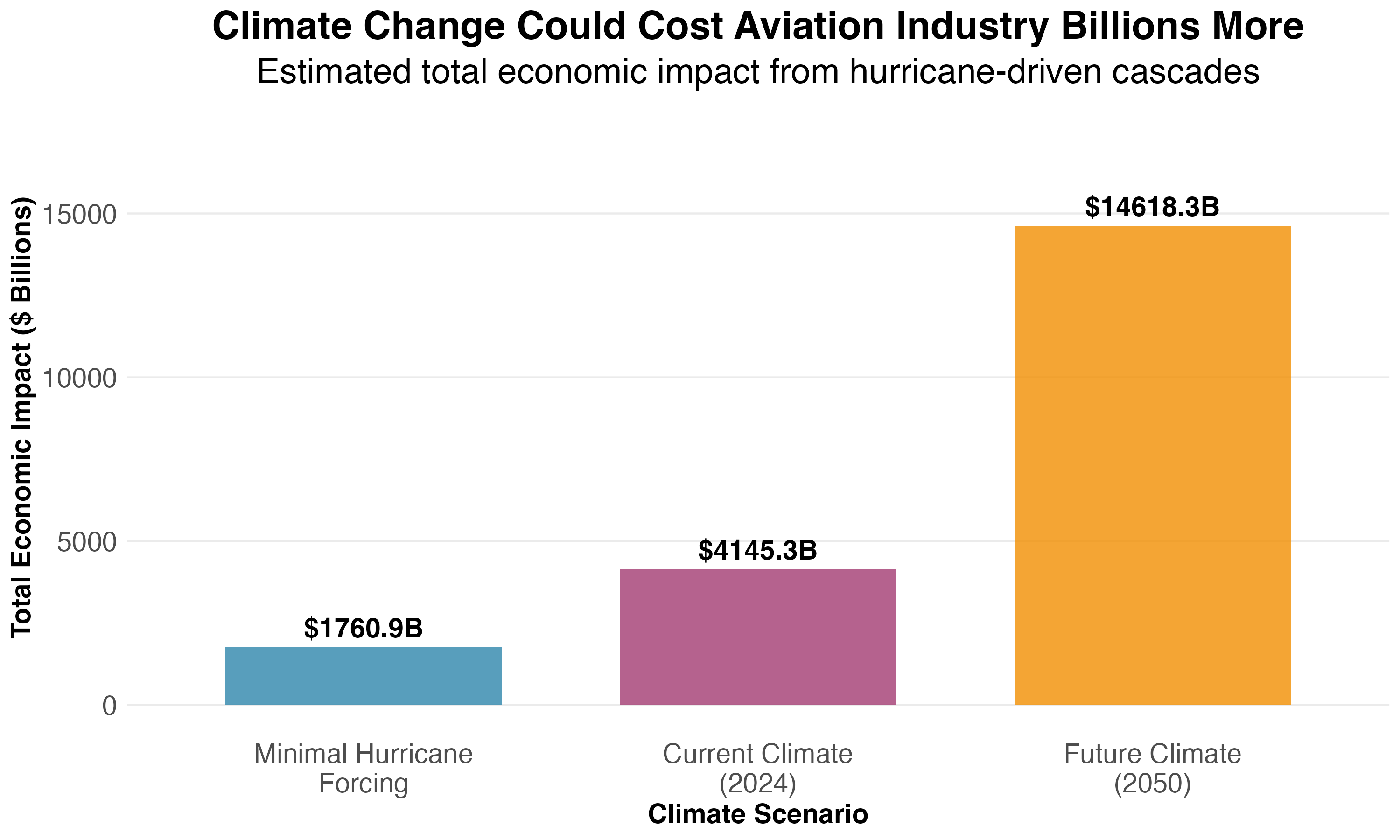}
   \caption{Economic losses increase 8.3×.}
   \label{fig:economic_cost}
\end{subfigure}
\caption{Climate-driven cascade amplification across four impact dimensions—event frequency, cascade severity, passenger disruption, and economic cost—under increasing hurricane intensity.}
\label{fig:soc_stress_testing}
\end{figure}

Simulations reveal that SOC systems exhibit nonlinear responses to forcing. Cascade frequency increases sharply with higher hurricane intensity. As shown in Figure~\ref{fig:cascade_frequency}, the rate of cascade events climbs from 119.9 under minimal forcing to 149.5 under current climate (+25\%) and 200.7 under future climate conditions (+67\%). This accelerating response highlights the network’s sensitivity to compounding stress and the onset of critical transitions.

Beyond frequency, the size distribution of cascading failures undergoes a regime shift. Cascades are categorized by operational impact: minor (1–2 airports), moderate (3–5), major (6–10), and catastrophic ($>$10). Figure~\ref{fig:cascade_severity} shows how increasing hurricane intensity shifts the system toward larger, more destructive events. Under future climate conditions, catastrophic cascades rise to 30.2\% of all events—2.4× more than under baseline forcing—while minor disruptions decline to 38.7\%.

To evaluate real-world implications, we simulate passenger disruptions by multiplying cascade size by a base disruption of 0.08 million passengers per airport. Figure~\ref{fig:passenger_impact} reveals dramatic increases in both median and extreme disruption events, with future scenarios producing frequent billion-passenger outages—indicating the potential for unprecedented transportation crises.

Economic impacts are estimated using the cost function:
\[
C_{\text{total}}(s) = s \cdot C_{\text{base}} + s^{1.2} \cdot C_{\text{cascade}},
\]
where \( C_{\text{base}} = \$15 \) million and \( C_{\text{cascade}} = \$10 \) million. This structure captures both linear per-airport damages and super-linear cascade amplification. As shown in Figure~\ref{fig:economic_cost}, total simulated impact escalates from \$1.76 trillion (minimal forcing) to \$14.6 trillion (future climate), representing an 8.3× increase driven by systemic instability—not just storm severity.

Together, these findings demonstrate that SOC-based stress testing offers actionable insights into nonlinear tipping points. A 67\% increase in event frequency under moderate climate forcing signals proximity to operational thresholds. The shift toward catastrophic dominance and 8.3× economic cost amplification reflect superlinear vulnerability growth—where failures become structurally inevitable, not just statistically more frequent.

Crucially, this framework allows exploration of adaptive strategies. Adjusting network parameters—such as increasing redundancy, implementing preemptive storm protocols, or targeting resilience investments—can be systematically tested within the SOC model. Unlike traditional risk assessments based on historical extrapolation, this approach accounts for emergent failure modes under future conditions. As a result, it provides policymakers with the scientific justification needed for forward-looking infrastructure planning, transforming climate risk management from reactive loss estimation to proactive systems-level optimization.

The SOC model enables identification of critical infrastructure nodes by mapping how climate change induces spatially heterogeneous risk across the airport network. We estimate the frequency of airport-specific cascade involvement as a function of hurricane exposure and climate forcing intensity:
\begin{equation}
V_{\text{airport}}(I_h) = \begin{cases}
10 \cdot I_h \cdot f_{\text{cascade}} & \text{if hurricane-prone} \\
2 \cdot I_h \cdot f_{\text{cascade}} & \text{otherwise}
\end{cases}
\end{equation}
where \(V_{\text{airport}}(I_h)\) denotes the expected number of cascade involvements, \(f_{\text{cascade}}\) is the system-wide cascade frequency, and the multipliers reflect elevated exposure and fragility in hurricane-prone coastal locations.

\begin{figure}[tb!]
\centering
\includegraphics[width=\textwidth]{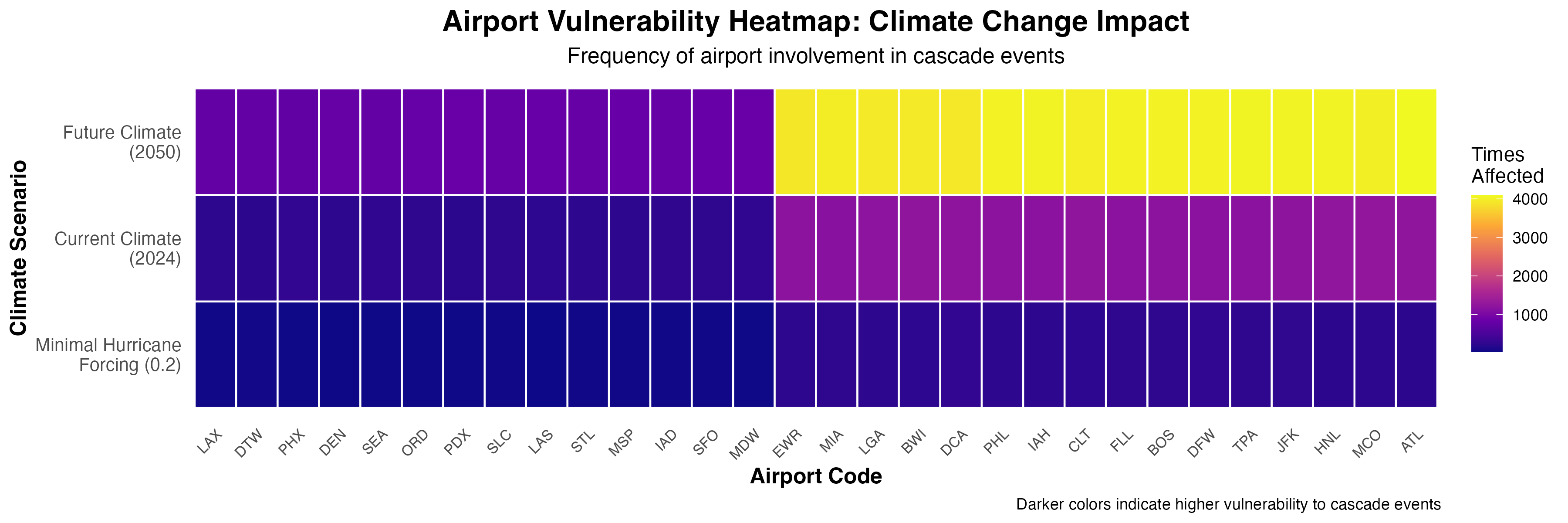}
\caption{Airport vulnerability heatmap showing frequency of involvement in cascade events across climate scenarios. Coastal hurricane-prone airports show sharply increasing vulnerability under future climate conditions. Darker colors indicate higher cascade involvement.}
\label{fig:vulnerability_heatmap}
\end{figure}

Figure~\ref{fig:vulnerability_heatmap} illustrates these spatial patterns, revealing a dramatic bifurcation between coastal and inland airports as climate intensity increases. Under future climate conditions, key coastal hubs—including ATL, MIA, JFK, MCO, TPA, and IAH—exhibit sharp increases in cascade involvement, transitioning from moderate to extreme vulnerability. In contrast, inland airports remain comparatively insulated, highlighting the spatial concentration of systemic risk.

This asymmetric vulnerability distribution arises from the dual role of hurricane-prone airports as both direct impact sites and cascade accelerators. Their geographic exposure to extreme weather, combined with their structural centrality in the network, compounds risk via failure propagation. The five-fold differential between coastal and inland nodes illustrates the emergent dynamics of criticality, where topological and climatological factors jointly determine system fragility.

These results underscore the value of geographically targeted resilience interventions. By reinforcing coastal hubs that dominate cascade pathways, decision-makers can achieve outsized reductions in systemic risk. The SOC model thus supports climate adaptation planning by identifying where mitigation resources yield the highest marginal benefit—at nodes where localized reinforcement produces system-wide stability gains.

\section{Conclusion and Future Directions} \label{sec:discussion}

This study demonstrates that modern aviation networks, optimized for efficiency, naturally evolve toward self-organized criticality (SOC)—a regime where small disruptions can trigger cascading failures with systemic consequences. This reflects the \textit{efficiency–resilience paradox}: design choices that improve day-to-day performance can undermine shock absorption, making highly efficient networks more fragile under stress.

We show that network structure and operational rules jointly determine the scale of disruptions. Dense interconnectivity and centralized hubs—often seen as strengths—can amplify failures, while decentralized or modular architectures contain them. Across all simulations, we observe universal power-law behavior, revealing that while specific cascades are unpredictable, the statistical structure of risk is not. This insight reframes disaster risk management: resilience depends less on rare shocks and more on latent structural fragility.

Our contribution is threefold: (1) an SOC model of the U.S. air traffic system that reveals how cascading failures emerge; (2) an early warning system that quantifies risk using power-law diagnostics and cascade dynamics; and (3) a climate stress-testing module that quantifies how extreme weather accelerates network collapse under future scenarios.

The integration of SOC theory, simulation, and real-time monitoring provides a roadmap for next-generation risk systems. By tracking early signals—such as shifting power-law exponents or clustered failures—decision-makers can anticipate tipping points and intervene before collapse. The approach applies beyond aviation: SOC explains fragility in power grids, hospitals, flood infrastructure, and supply chains—all networks where compounding stress and tight coupling drive systemic risk.

Ultimately, SOC enables a new paradigm for resilience. The goal is not to prevent all failures, but to understand how risk emerges, where it concentrates, and how systems can be made robust—not just to known threats, but to the unpredictable dynamics of an interconnected world.

\bibliographystyle{cas-model2-names}

\bibliography{cas-refs}






\bio{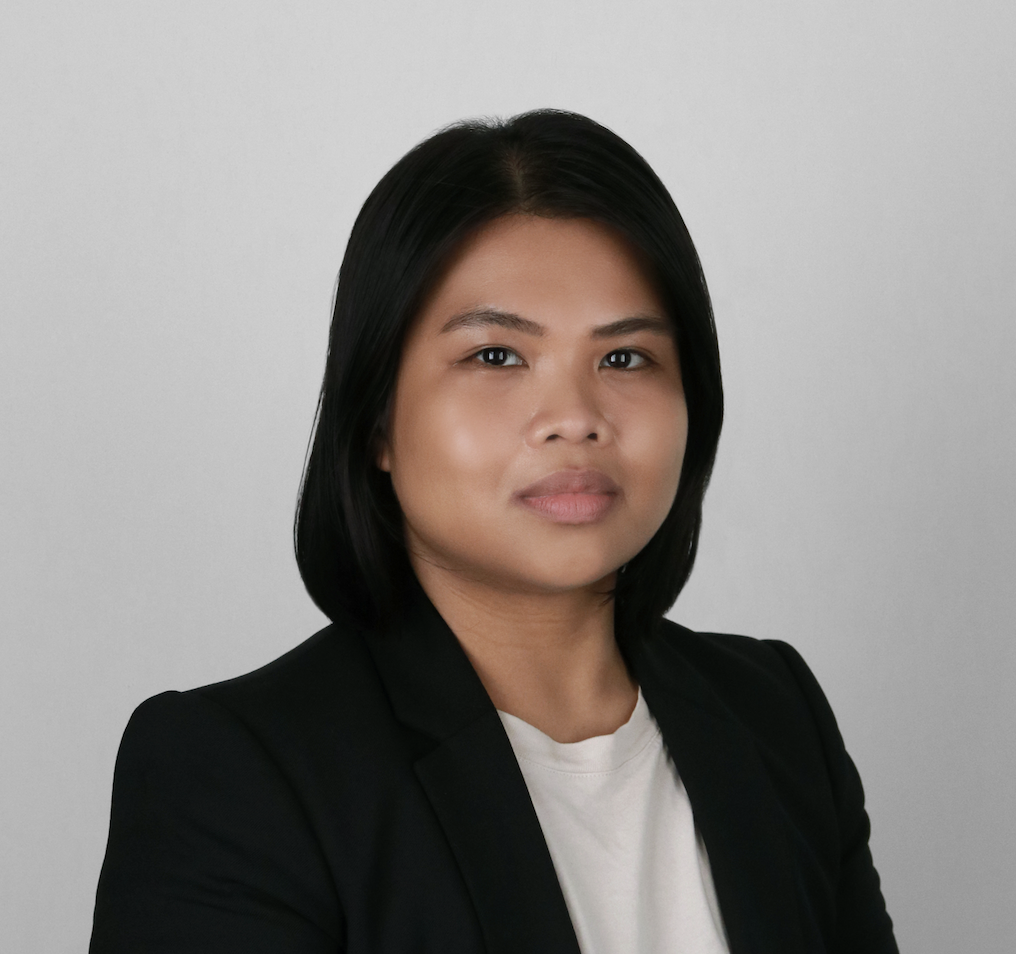}
Mary Lai O. Salva\~na is an Assistant Professor of Statistics at the University of Connecticut (UConn). Prior to joining UConn, she was a Postdoctoral Fellow at the Department of Mathematics at University of Houston. She received her Ph.D. in Statistics at the King Abdullah University of Science and Technology (KAUST), Saudi Arabia. She obtained her BS and MS degrees in Applied Mathematics from Ateneo de Manila University, Philippines, in 2015 and 2016, respectively. Her research interests include extreme and catastrophic events, risks, disasters, spatial and spatio-temporal statistics, environmental statistics, computational statistics, large-scale data science, and high-performance computing.
\endbio

\bio{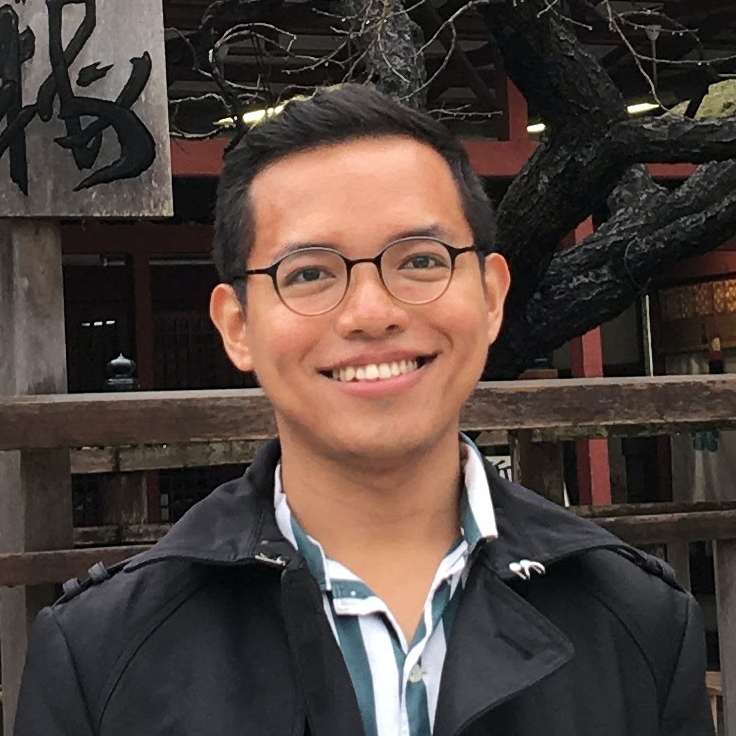}
Harold Jay M. Bolingot holds a Bachelor of Science in Computer Engineering from Ateneo de Manila University, Philippines, and a Master of Engineering in Assistive Robotics from Kyushu Institute of Technology, Japan. His research expertise lies in machine learning, with a focus on convolutional neural networks, computer vision, and biomedical signal processing.\vspace{1cm}
\endbio

\bio{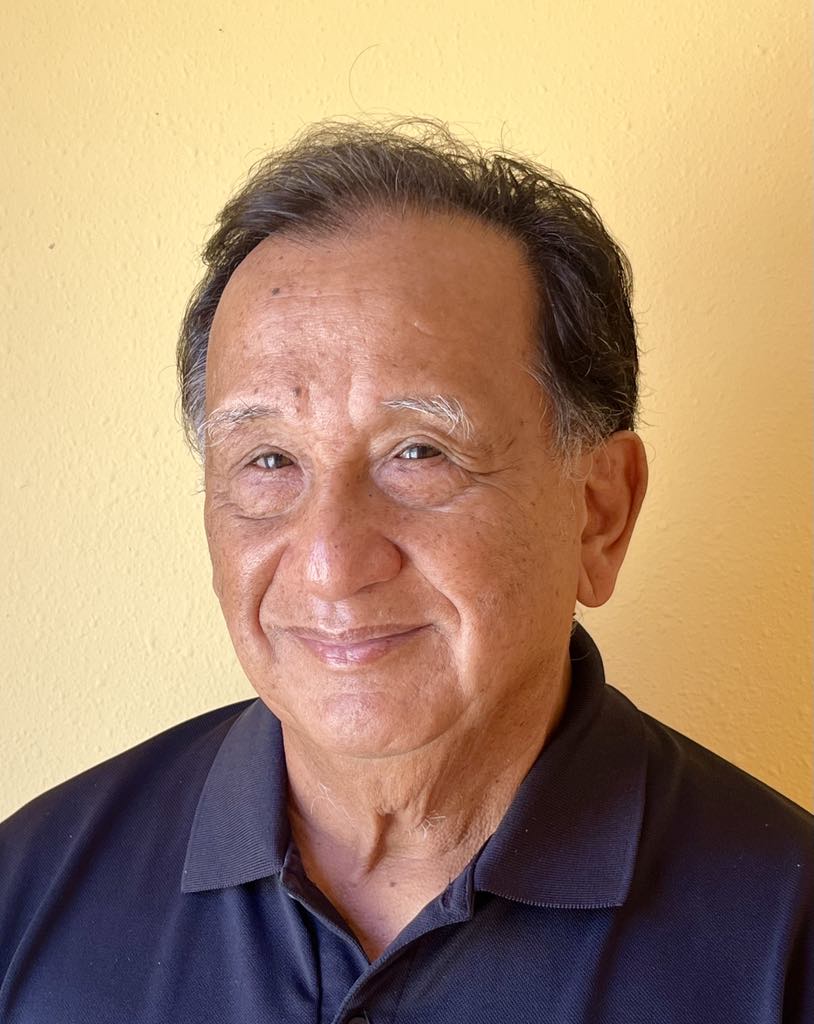}
Gregory L. Tangonan graduated in Physics from Ateneo de Manila University and earned his PhD in Applied Physics from the California Institute of Technology. He spent 32 years at Hughes Research Laboratories in Malibu, California, retiring as Lab Director of the Communications Laboratory. His expertise spans fiber optics, wireless communications, materials science, and complex systems science. He has published over 200 papers and holds 49 U.S. patents. After retiring, he returned to his alma mater and founded the Ateneo Innovation Center (AIC), which is actively engaged in disaster risk and resilience research. AIC has developed Resilience Hubs for nationwide deployment in the Philippines and has been at the forefront of research on cascaded disasters and risk assessment.
\endbio

\end{document}